# Dynamically controlled plasmonic nano-antenna phased array utilizing vanadium dioxide

Gregory Kaplan[1], Koray Aydin[2] and Jacob Scheuer[1,*]

[1]*Department of Physical Electronics, School of Electrical Engineering, Tel Aviv University, Tel Aviv 69978, Israel*
[2]*Department of Electrical Engineering and Computer Science, Northwestern University, Evanston, IL, 60208 USA*
*\*kobys@eng.tau.ac.il*

**Abstract:** We propose and analyze theoretically an approach for realizing a tunable optical phased-array antenna utilizing the properties of $VO_2$ for electronic beam steering applications in the near-IR spectral range. The device is based on a 1D array of slot nano-antennas engraved in a thin Au film grown over $VO_2$ layer. The tuning is obtained by inducing a temperature gradient over the device, which changes the refractive index of the $VO_2$, and hence modifies the phase response of the elements comprising the array, by producing a thermal gradient within the underlying PCM layer. Using a 10-element array, we show that an incident beam can be steered up to $\pm 22°$ with respect to the normal, by applying a gradient of less than 10°C.

## 1. Introduction

The physics and applications of plasmonic nano-antennas at optical frequencies have been the focus of numerous studies during the last decade. These studies have resulted in applications in diverse fields ranging from spectroscopy [1] and near-field microscopy [1,2] to non-linear optics [3], holography [4,5], sensing [6] and many others. As the spectral properties of such antennas are determined by their geometry and surroundings, much research was carried out to determine the impact of these parameters on the resonance frequency, bandwidth, efficiency and radiation profile of the nano-antennas [7,8]. These studies have resulted in a multitude of possible nano-antenna geometries, such as dipole antennas [1,9,10], slots [6,11], bow-ties [1,2,6], split ring resonators [10] and Vivaldi antennas [7] to name a few. In addition to geometry, the choice of materials (those composing the antenna and the surroundings) has also a strong impact on device properties, leading to the study of a wide variety of potential metallic and dielectric materials.

The high sensitivity of nano-antenna parameters to their dimensions and surroundings [13], suggests that for practical applications it is highly desired to have the ability to tune the response (particularly the resonance frequency) post-fabrication. Such ability is not only important for compensating for fabrication errors, but also for obtaining devices with larger functionality. Consequently, developing such capability is the subject of ongoing research with numerous approaches that are actively investigated [13–17]. Since the resonance frequency of a nano-antenna is highly affected by its dielectric environment [13], most efforts were focused on approaches which modify the refractive index of nano-antenna's immediate surroundings. Among the proposed methods are liquid crystals [15], elastomeric stretching [16], free-carrier effects [17], as well as phase change materials [18] and nonlinear effects [14].

In parallel to tuning the resonance frequency of a nano-antenna it is also possible to tune the phase response of the antenna at a given frequency. The immediate application of this capability is the realization of optical phased arrays. Similar to their radio frequency (RF) counterparts, optical phased arrays enable beam steering and shaping without mechanical motion. This opens up a variety of applications ranging from laser steering for free-space optical communication [19] and LIDARs to dynamic holographic displays [4] and chip-scale photonic devices [20]. While RF phased arrays are generally large, the optical counterparts can be sufficiently small for on-chip integration, enabling all the above applications while occupying a very small footprint. Furthermore, compared to existing mechanical solutions for optical beam steering, phased array devices can reduce considerably weight and power consumption. Consequently, several approaches for the realization of such devices have been studied and demonstrated, exhibiting varying degrees of steering capability and fabrication complexity.

Doylend et al. [21] presented an array of dielectric gratings which are fed through on-chip phase shifters. Shortly after that, a large scale device based on a similar concept was demonstrated by Sun et al. [22]. A similar tunable phased-array device in which the output gratings were replaced by metallic nano-antennas was recently demonstrated as well [23]. This approach essentially transfers the RF phased array concept to the optical domain, enabling $2\pi$ phase shift at each pixel. However, due to the need for individual feed waveguide for each pixel, the method suffers from high manufacturing complexity and relatively modest efficiency. In addition, waveguide based phase shifters are relatively large, thus limiting the minimum size of the single antenna cell. Consequently, despite the $2\pi$ phase shift capability, obtaining a large steering angle is a challenging task (e.g. a steering angle of 8° was demonstrated in [23]).

Some of these difficulties can be overcome by using a reflectarray configuration, which can provide a wide steering angle and high efficiency [4,24–33]. Recently, a device utilizing such approach for beam switching applications was theoretically studied by Zou et al. [20], where GeSbTe, a high-temperature phase change material, was used for controlling the phase response of the array.

In this paper we propose and analyze a reflective nano-antenna array utilizing vanadium dioxide ($VO_2$) as a means of continuous electrical steering of an incident infrared beam. $VO_2$ is a well-known phase-change material (PCM) which exhibits a semiconductor-to-metal phase transition around 67°C [34]. This phase transition is accompanied by a large shift in the dielectric index of the material [35]. Consequently, and due to its picosecond scale transition time [36] numerous $VO_2$ based applications have been proposed . Among these are optical memory cells [37], photonic and plasmonic modulators [38], RF switches in the microwave range [39] and fast-switching mirrors for laser pulse shaping [40]. Another useful property compared to other PCMs (e.g. GeSbTe and AgInSbTe) is that the phase transition of $VO_2$ occurs relatively close to room temperature, rendering it highly suitable for electrical control (e.g. by using resistive heating).

The possibility of nano-antenna resonance tuning using $VO_2$ has been presented and discussed in several studies. Lei et al. [41] observed a blueshift of 30 nm at a wavelength of 650 nm in the resonance peak of 100 nm Au spheres positioned on a $VO_2$ substrate when heating the device from 35° to 90°. Using an array of silver dipole antennas, a peak reflectance shift of up to 110nm was demonstrated in a similar setup [42]. At longer wavelengths, resonance shift exceeding 1μm at $\lambda \approx 10\mu m$ was demonstrated in [43] and 100 nm shift was observed for split ring resonator metasurfaces on $VO_2$ at the telecom wavelength range [35]. Nevertheless, to the best of our knowledge, the possibility of utilizing PCM for continuous beam steering/shaping has not been proposed and studied. Such steering can be obtained due to the shift of the phase response of a nano-antenna which is accompanied with a shift of its resonance frequency.

Fig. 1 depicts the proposed device consisting of a 1D finite array of slots in a gold film which is deposited over a layer of $VO_2$, supported by an alumina substrate. By inducing a temperature gradient in the $VO_2$ film, a corresponding refractive index shift is generated, and the resonance wavelength of each nano-antenna is shifted. When an incident monochromatic beam impinges upon the array, the shift in the resonance wavelengths also induces a shift in the phases of the reflected wave from each nano-antenna. This progressive phase shift enables a direct control of the angle of the reflected beam, similar to a broadside phased array in the RF frequency range.

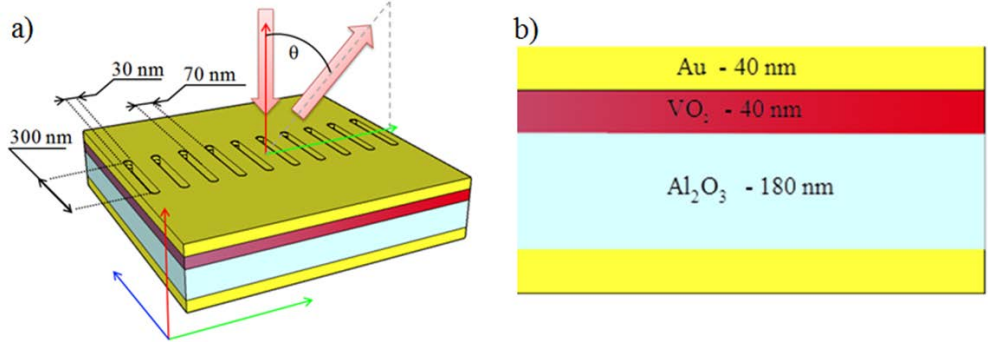

Fig. 1. a) Schematic of the device; thermal gradient is to be applied along the direction of the green axis b) Layer structure.

Phased arrays in the microwave frequency is a well-known concept which is employed, among other examples, in communication links, radars, and radio astronomy. By controlling the phase of each element in an array of many such elements, it is possible to almost arbitrarily shape the antenna radiation profile [44]. Specifically, if the far-field amplitude and phase response of a single element in the array is given by the $D(\theta,\varphi)$, and the antennas are placed at points $\vec{r}_i$ and fed with a signal $A_i \exp(j\phi_i)$, where $A_i$ and $\phi_i$ are respectively the amplitude and phase at each element, then the far field pattern is given by :

$$D_{arr}(\theta,\varphi) = F\left\{\sum_i A_i \exp(j\phi_i)\delta(\vec{r}-\vec{r}_i)\right\} \times D(\theta,\varphi) \quad (1)$$

where $F\{\ \}$ indicates the spatial Fourier transform.

Eq. (1) indicates that overall radiation pattern consists of two contributions – the element factor and the array factor, where the later can be used for steering and shaping the far field amplitude distribution. A linear progressive phase shift in the form of $\phi_i = n\Delta\phi$ would steer the main lobe by an angle $\theta$ according to equation (2), where $\lambda$ is the incident wavelength, and $d$ the array spacing.

$$\theta = \arcsin\left(\frac{\Delta\phi}{2\pi}\cdot\frac{\lambda}{d}\right) \quad (2)=$$

In order to induce a progressive phase shift between adjacent nano-antennas, we propose to use the disorderly manner in which VO2 undergoes a phase transition, depending on the crystalline grain composition of the film. Qazilbash et al. [45] showed that a thin $VO_2$ film grown over an $Al_2O_3$ substrate has a multidomain structure, where the typical grain size is smaller than 50 nm. In the same work, a temperature dependence of the fraction of insulating/metal $VO_2$ was shown to be a nearly linear between 62° C and 68° C. Therefore, it is possible to use a thermal gradient between these extreme temperatures which is induced over the length of the array to change the refractive index at the nano-antenna sites. To model the refractive index of the $VO_2$ film as a function of the temperature we used the data of [35] for the hot and cold phases and interpolated for the intermediate stage using a model described in Section 2.

The rest of this paper is structured as follows: Section 2 describes the material model we used for $VO_2$ thin film. In Section 3 the choice of geometry is explained. Finally, in Section 4 we present the essential results of this work, together with an accurate description of the numerical simulation used to establish them. Section 5 serves as the conclusion and outlines the implications and future possibilities of the obtained results.

## 2. Optical properties of VO$_2$ in the transition region

At low temperatures, VO$_2$ exhibits semiconductor properties, with sufficiently low conductivity to be considered an insulator. This low conductivity results in relatively low (although non-negligible) losses in the optical and near-IR frequency ranges. The refractive index varies greatly in this range, from approximately 2.5 at 500 nm to almost 4 at 2 μm, as shown in Fig. 2. The refractive index properties change dramatically when the material undergoes a phase transition to a metallic phase (labeled 'Hot phase' in Fig. 2) [35]. The absorption increases significantly, growing larger for longer wavelengths. Correspondingly, the refractive index changes, with a large shift occurring in the visible to near IR range below 1500 nm and in the far IR.

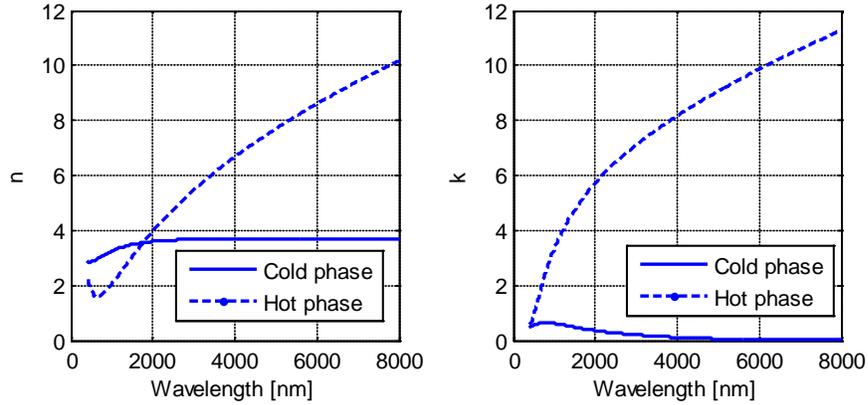

Fig. 2. Refractive index data for VO$_2$ [35]

The transition of VO$_2$ from insulating to metallic phase is caused by a structural change from a monoclinic to a rutile (tetragonal) lattice structure [46]. This change can be triggered by either strain, electric field or strong optical excitation, but most typically by increasing the temperature. The exact mechanism of the transition was long debated, with both Coulomb repulsion (Mott transition, [47]) and charge density waves (Peierls transition) proposed as a possible model, until recently, when Budai et al. [48] showed that the phase transition can be completely calculated from basic principles using Peierls instability coupled with an anharmonic phonon contribution. In a single VO$_2$ crystal , this transition occurs abruptly over the entire crystal domain once a critical temperature or strain is reached [49]. Clearly, such characteristic is not suitable for obtaining a gradual change in refractive index.

Unlike intrinsic crystalline vanadium dioxide, thin VO$_2$ films consist typically of many nano-crystal domains, or nanograins. Using a scanning probe microscope, Qazilbash et al. [47] have shown that the transition in such films at the nanoscale occurs gradually, with different grains undergoing phase transition at different temperatures, thus forming a mixture of the two phases at intermediate temperatures. The temperature vs metallic phase fraction relationship was measured by the authors of [45] for an 80 nm VO$_2$ film over alumina substrate and is shown in Fig. 3 below. Clearly, dependence of the metallic phase content of a multidomain thin VO$_2$ film on the temperature can be approximated by a linear relation. As the typical size of the grains is much smaller than the wavelength, such mixture is expected to produce a relatively smooth variation of the optical properties within a small temperature range, thus rendering VO$_2$ films attractive as medium for controlling the refractive index in the vicinity of nano-antennas.

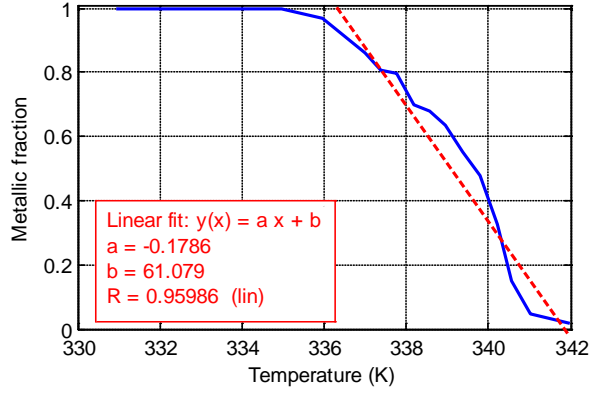

Fig. 3. Metallic fraction of $VO_2$ vs temperature for an 80nm film over alumina (adapted from [45]) with a linear regression analysis overlay

In order to calculate the impact of the mid-transition $VO_2$ on the optical response of a nano-antenna it is required to develop a model for the relation between the refractive index of the film and the metallic/semiconducting mixture (or equivalently, the temperature). In view of the thin films multidomain formation in the transition region between the two phases, it is reasonable to expect that the corresponding (n, k) parameters would be between those of the semiconducting and metal phases. An experimental evidence to such behavior is provided by a study of the temperature dependent electrical resistance of $VO_2$ nanorods [50]. When such nanorods are attached to a substrate, they exhibit a 1-D grain structure, with grain boundaries along the length of the rod. The study in [50] showed that when the temperature was modified (in the transition range), the measured resistance of the nanorods varied continuously between the extreme values (metallic and semiconducting).

We employ a simple Lorentz-Lorenz model to evaluate the refractive index and absorption coefficient for an arbitrary metallic phase fraction. This model is an adaptation of the more general effective medium theory [51] for the case of binary mixture of isotropic linear dielectrics.

For a mixture of two linear dielectrics, the polarization density $P_{mix}$ of the mixture is the weighted sum of individual polarizations $P_1, P_2$:

$$P_{mix} = \phi P_1 + (1-\phi) P_2 \qquad (3)$$

where $\phi$ is the volume fraction of the material with polarization $P_1$. Using the Clausius-Mossotti relation, we can replace polarizability by a dielectric constant, writing

$$\frac{\varepsilon_{mix}-1}{\varepsilon_{mix}+2} = \phi \frac{\varepsilon_1-1}{\varepsilon_1+2} + (1-\phi) \frac{\varepsilon_2-1}{\varepsilon_2+2} \qquad (4)$$

Strictly speaking, Eqs. (3) and (4) are only valid for isotropic materials while $VO_2$ exhibits anisotropy due to its crystal structure (monoclinic or tetragonal). However, in our proposed device, only a single polarization is excited, due to the properties of the impinging illumination. Thus, the anisotropy of the $VO_2$ is not important for the studied scenario and the model should be considered valid, with $P_{1i}$ and $P_{2i}$ used in Eq. (3) instead of $P_1, P_2$, where $i$ is the polarization direction. For the case of multi-domain thin films with arbitrary grain orientations, only the average response can be observed. Fig. 4 depicts the calculated $n$ and $k$ for various mixtures of cold and hot $VO_2$ according to Eq. (4). The curves for 0% and 100% cold phase are identical to those of Fig. 2, based on [35].

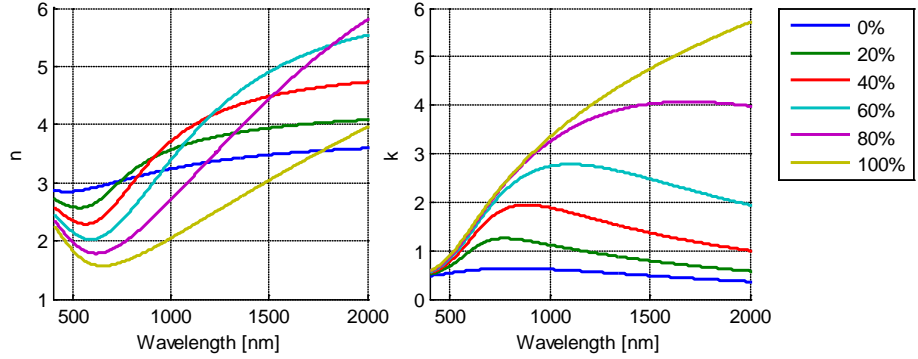

Fig. 4. Simulated refractive index for $VO_2$ undergoing a phase transition, with varying hot/cold phase content ratio; Percentages are for the cold fraction.

Combining the results of the Lorentz-Lorentz model for the dielectric coefficients of mid-transition $VO_2$ with the relation between the metallic fraction in the temperature (Fig. 3), it is straightforward to obtain a relation between the temperature and the refractive index of the $VO_2$ film. In turn, this relation can be used for determining the impact of the temperature gradient on the steering angle.

It is important to note that despite the stochastic nature of the multi-domain phase transition process, the relationships depicted in Fig. 4 are expected to be stable and repeatable. Qazilbash et al. [45] showed that reported repeated measurements of the thin $VO_2$ film at a constant temperature result in identical patterns of the coexisting phases and, hence, constant optical properties for that temperature. Furthermore, it was shown in [50] that multidomain $VO_2$ crystals on a substrate cycled repeatedly across a phase transition exhibited a constant resistance vs temperature curve, suggesting similar behavior for the refractive index.

## 3. Device design and optimization

For beam steering purposes, the simplest device structure is a 1D optical nano-antenna array consisting of equispaced elements (as shown in Fig. 1), similar to the RF broadside antenna array configuration. For an array comprising a fixed number of elements, smaller spacing yields a wider aperture main lobe, while facilitating a larger steering angle [44]. A key factor in the design of a tunable reflectarray is the choice of a radiating element (i.e. the antenna). Ideally, the antenna should exhibit high radiation efficiency at the desired wavelength and low Ohmic losses. In addition, the spectral response of the individual elements should be sensitive to variations of the substrate refractive index as this is the mechanism utilized for tuning the nano-antenna resonance frequency (and phase response). Finally, the elements should be simple to fabricate using conventional lithography methods.

A narrow resonance lineshape is generally accompanied by a high phase to frequency slope at the vicinity of resonant wavelength. Consequently, in order to obtain high sensitivity to refractive index changes (i.e. large $\Delta\phi/\Delta n$ where $\phi$ and $n$ are respectively the antenna phase response and the substrate refractive index), it is desired to use antennas exhibiting a narrow linewidth. It is well known that the spectral width of plasmonic resonance is inversely correlated to the aspect ratio of a nano-antenna [6,52]. Therefore, it is advantageous to employ narrow antennas such as dipoles and nano-slots as the elements comprising the array. Here, we focus on slot antennas, primarily because narrow slots are simpler to fabricate (e.g. using focus ion beam milling) than narrow dipoles [53]. We optimized the individual nano-antennas for operation at 1064 nm because of the various relevant applications for dynamic beam steering at that wavelength. The optimized element consists of 30 nm wide slots which are etched into a 40 nm thick Au layer. The length of the slots is set to 300 nm, which yields the maximal $\Delta\phi/\Delta n$ of 1.97 rad/RIU (RIU - refractive index unit) at the design wavelength (1064 nm). The spacing between adjacent nano-slot centers is 100 nm.

The array is positioned on a 40 nm thick $VO_2$ layer. This thickness is a compromise between two competing effects – a thicker film has larger impact on the plasmonic resonance shift, while a thinner film consists of smaller grains and so has a smoother refractive index variation with temperature. Although it may seem that placing the PCM inside the slot would provide maximum sensitivity of the antenna phase response to the material phase transition, the metallic nature of hot-phase $VO_2$ results in high absorption in the slot and hence low radiation efficiency of the "hot" array elements. Finally, the $VO_2$ layer is grown on top of a $\lambda/4$ thick alumina ($Al_2O_3$) substrate [35,45] with an Au backplane which enhances the overall radiation efficiency of the array [4,32]. The thickness of this layer is of lesser significance, as long as it is sufficiently thick to reflect all of the incident radiation.

Fig. 5 depicts the amplitude and phase response of an array of identical elements for incident wavelength ranging between 0.6-1.6 µm. The response was calculated using a commercial Finite Difference Time Domain (FDTD) simulation tool (see Section 4) where the dielectric properties of the $VO_2$ layer were set according to the plots in Fig. 4. The results can be readily understood as redshifting and broadening of the antenna resonance due to lower refractive index and the higher absorption of the $VO_2$ at higher temperatures. As the metallic fraction of the substrate increases, the phase response of a single element at 1064 nm undergoes a phase shift of ~2.2 rad while exhibiting a nearly flat amplitude response. For the presented configuration, such phase shift tuning can be obtained from 1000 nm to 1200 nm.

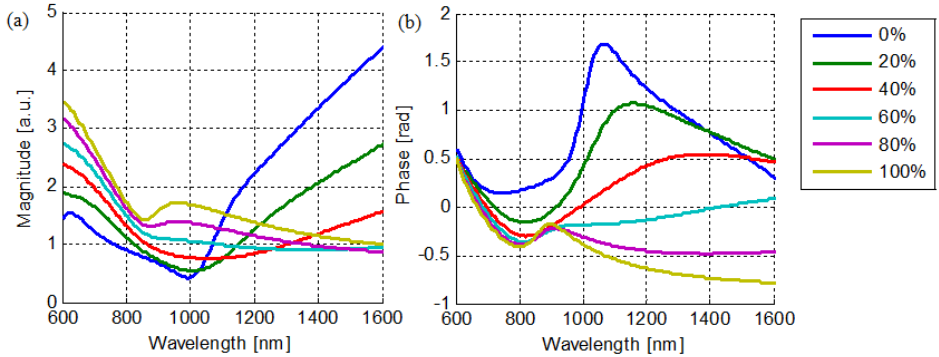

Fig. 5. The amplitude (a) and phase (b) response of a an array of identical slot nano-antenna elements for varying metallic phase fraction of $VO_2$ thin film

Assuming, for example, an array consisting of 10 element array with interelement spacing of 100 nm, then based on a maximum phase shift of 2.2 rad between the extreme elements in the array the maximal steering angle is approximately 22°. This is also shown in Section 4 below where this result is obtained for the full device.

## 4. Results

The scattering of an incident beam at 1064 nm from the proposed device was calculated using a commercial FDTD tool (FDTD Solutions 8.6.4, Lumerical Inc.). The dimensions of the calculation space were 1.5x1x1.5 µm with spatial resolution of 40nm, where fine resolution (2nm in each dimension) was used at the antenna sites. Perfectly matched layer (PML) boundary conditions were employed at the edges of the simulation region. The device is excited by a linearly polarized plane wave which impinges at normal incidence with respect to the metallic layer. The polarization was chosen to excite the dipole moment of the slots, that is the E-field perpendicular to their long axis. The material properties of the gold and alumina were obtained from the CRC Handbook of Chemistry and Physics [54] and Palik's Handbook of Optical Constants [55] respectively, while the $VO_2$ properties were based on the model described in Section 2.

Fig. 6 depicts the amplitude and phase of electric distribution at the center of nano-slots array where maximal thermal gradient was applied across the array (i.e. completely semiconducting $VO_2$ at the leftmost and completely metallic $VO_2$ at the rightmost nano-antennas). As seen in the figure, the electric field is focused in the nano-slots (as can be expected) and a continuously decreasing phase profile is formed between the leftmost and rightmost elements. This phase gradient corresponds to the changes in metallic/semiconducting phases ratio which is induced by the temperature gradient as discussed in Section 3. We also note that the intensities inside the individual nano-antennas are not equal. This is attributed to higher losses of the $VO_2$ in the metallic phase (see Fig. 4). Although this intensity variation may reduce the maximal attainable steering angle and slightly distort the beam profile, its impact is rather minor as shown in Fig. 7 and Fig. 8.

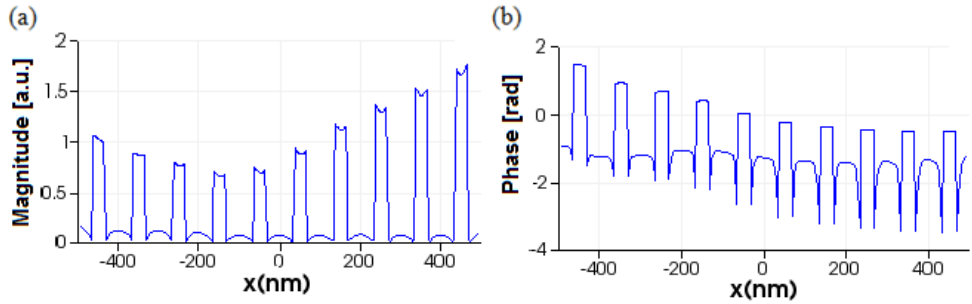

Fig. 6. The amplitude (a) and phase (b) of the local field along the center of the nano-antennas array for a maximal thermal gradient along the array axis

Fig. 7 shows the far-field distributions of the beam scattered from the steerable reflectarray for three different temperature gradients. The leftmost panel depicts the emitted beam for a uniform $VO_2$ temperature profile, while the other two plots show the beam steering caused by increasing the thermal gradient across the array. An approximately ±22° sweep angle is obtained for the case of maximum temperature slope (i.e. pure metallic/semiconducting phases at both sides), which is in excellent agreement with the results of Section 3. Setting the thermal gradient to a fraction of the maximum value (i.e. mixed phases in the rightmost element), yields intermediate steering angles, as shown in the center panel of Fig. 7 (steering of 12°).

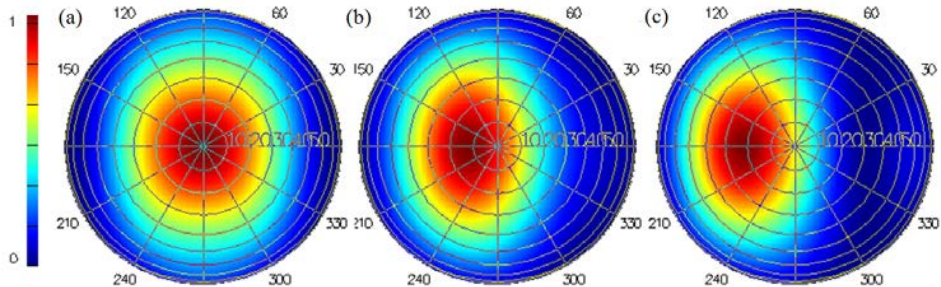

Fig. 7. Far field plots of $|E|^2$ of the beam scattered from the array, with thermal gradient across the array of (a) 0%, (b) 33% and (c) 100% of maximum

Fig. 8 depicts an angular cross-section of the far field distribution over $\theta = -90...90°$ for various temperature gradients across the device. The angular response exhibits a single emission lobe which is ~45° wide (FWHM) and a steering angle that can be continuously varied over a range of 44°. The intensity of the reflected beam depends on the steering angle due to rising reflectivity with increasing hot phase fraction of $VO_2$, as shown in Fig. 5 (a) for a single nano-antenna. The intensity variation is in the order of ~50% where overall reflectivity increases for larger fractions of the metallic phase. Nonetheless, this dependence can be easily compensated by varying the intensity of incident beam according to the desired steering angle.

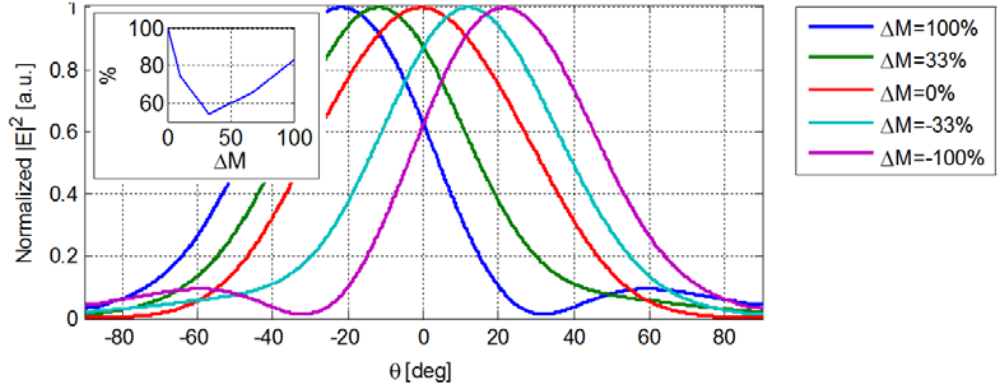

Fig. 8. A $\varphi = 0$ angular plot of the far field response of the proposed device for varying difference in metallic fraction content of $VO_2$ between the two ends of the array (ΔM). Inset: the dependence of the amplitude of the reflected beam on ΔM

**5. Discussion and conclusions**

We presented an approach for realizing dynamic beam steering reflectarray at the near IR range by using $VO_2$ as a means for controlling the phase of individual slot nano-antennas. By generating a thermal gradient in the antenna substrate, we induce a progressive phase shift in the scattering response of the individual elements comprising the array, and hence obtaining steering of the reflected beam. Using 3D FDTD simulations we presented the ability to tune the phase response over a range of 2.2 rad, yielding a steering angle of ±22° from the normal. Obtaining larger steering angles should also be possible by further optimizing the geometric parameters of the device, especially if non-identical antenna elements are employed. For the device configuration under study, a bandwidth of about 20% around the central wavelength can be expected without substantial degradation of the far-field response.

It should be noted that some fraction of the incident radiation might be absorbed by the device and hence increase its temperature. This is in addition to the bias heating used for inducing the phase change. Since the optimal operating temperature is approximately 30°C above room temperature, the impact of this excess heating caused by the impinging beam can be easily compensated by reducing heating bias.

An important parameter of the proposed device is its maximal switching time (or equivalently – steering rate). Although thin $VO_2$ films have been shown [36] to undergo a full phase transition on the order of several picoseconds, the actual switching time of the device is determined primarily by its heat capacity and heater implementation. As these details are realization dependent, they are beyond the scope of this work and will be addressed in future studies.

While the presented study was focused on beam steering, additional applications utilizing the same mechanism can be easily envisioned, especially if employed for controlling the phase response of the individual nano-antennas. The relatively low phase transition temperature of $VO_2$ and its picosecond transition time render this approach highly suitable for electrical control of plasmonic nano-antennas. By controlling the temperature of the individual nano-antennas (e.g. by resistive heating), it is possible to induce an arbitrary phase profile on the reflected beam, leading to a wide variety of applications such as real time holography, tunable optical components, etc.

In conclusion, we have shown the possibility of continuous beam steering using a nano-antenna array configuration utilizing a mid-transition, near room temperature, phase change material. This approach has great potential for realizing electrically controlled reflectarrays and can open new avenues in the field of plasmonic on-chip devices for subwavelength optical phase control, such as electrically controlled holography, free-space and on-chip optical communication, LIDAR, imaging and many more.

## Acknowledgments

This program was financed by Pearls of Wisdom, an Israeli registered non-profit association.